\documentclass{aa}

\usepackage{lineno}

\usepackage{graphicx}
\usepackage[varg]{txfonts}
\usepackage{natbib}
\bibpunct{(}{)}{;}{a}{}{,}
\usepackage{array}
\usepackage{xspace}
\usepackage{lscape}
\usepackage{url}

\usepackage[hyperfootnotes=false, linktocpage=true, breaklinks=true, colorlinks=true, linkcolor=blue, citecolor=blue, urlcolor=blue]{hyperref}
\usepackage[all]{hypcap}
\usepackage{longtable}
\usepackage{afterpage}
\usepackage[dvipsnames]{xcolor}
\usepackage{xfrac}

\usepackage{placeins} 

\newcommand{\FeKa}{Fe K\ensuremath{\alpha}\xspace}
\newcommand{\FeKb}{Fe K\ensuremath{\beta}\xspace}
\newcommand{\civ}{\ion{C}{iv}\xspace}
\newcommand{\fexxv}{\ion{Fe}{xxv}\xspace}
\newcommand{\fexxvi}{\ion{Fe}{xxvi}\xspace}
\newcommand{\kms}{\ensuremath{\mathrm{km\ s^{-1}}}\xspace}
\newcommand{\NH}{\ensuremath{N_{\mathrm{H}}}\xspace}
\newcommand{\vout}{\ensuremath{v_{\mathrm{out}}}\xspace}
\newcommand{\sigv}{\ensuremath{\sigma_{v}}\xspace}

\newcommand{\xabs}{\xspace{\tt xabs}\xspace}
\newcommand{\delt}{\xspace{\tt delt}\xspace}
\newcommand{\vgau}{\xspace{\tt vgau}\xspace}
\newcommand{\pow}{\xspace{\tt pow}\xspace}

\newcommand{\nustar}{{\it NuSTAR}\xspace}
\newcommand{\xmm}{{\it XMM-Newton}\xspace}
\newcommand{\xrism}{{XRISM}\xspace}
\newcommand{\chandra}{{\it Chandra}\xspace}
\newcommand{\swift}{{\it Swift}\xspace}

\newcommand{\ergflux}{{\ensuremath{\rm{erg\ cm}^{-2}\ \rm{s}^{-1}}}\xspace}

\newcommand{\cm}{{\ensuremath{\rm{cm}^{-2}}}\xspace}
\newcommand{\spex}{\xspace{\tt SPEX}\xspace}
\newcommand{\pion}{\xspace{\tt pion}\xspace}

\newcommand{\logxi}{\ensuremath{{\log \xi}}\xspace}

\newcommand{\lya}{Ly\ensuremath{\alpha}\xspace}
\newcommand{\lyb}{Ly\ensuremath{\beta}\xspace}

\newcommand{\ngc}{{NGC~3783}\xspace}

\mathchardef\mhyphen="2D

\begin{document}

\title{Delving into the depths of NGC 3783 with XRISM}
\subtitle{I. Kinematic and ionization structure of the highly ionized outflows}

\author{
Missagh Mehdipour \inst{1}
\and
Jelle S. Kaastra \inst{2,3}
\and
Megan E. Eckart \inst{4}
\and
Liyi Gu \inst{2,3}
\and
Ralf Ballhausen \inst{5,6,7}
\and
Ehud Behar \inst{8,9}
\and
Camille M. Diez \inst{10}
\and
Keigo Fukumura \inst{11}
\and
Matteo Guainazzi \inst{12}
\and
Kouichi Hagino \inst{13}
\and
Timothy R. Kallman \inst{6}
\and
Erin Kara \inst{9}
\and
Chen Li \inst{3,2}
\and
Jon M. Miller \inst{14}
\and
Misaki Mizumoto \inst{15}
\and
Hirofumi Noda \inst{16}
\and
Shoji Ogawa \inst{17}
\and
Christos Panagiotou \inst{9}
\and
Atsushi Tanimoto \inst{18}
\and
Keqin Zhao \inst{3,2}
}

\institute{
Space Telescope Science Institute, 3700 San Martin Drive, Baltimore, MD 21218, USA \\ \email{missagh.mehdipour@gmail.com}
\and
SRON Netherlands Institute for Space Research, Niels Bohrweg 4, 2333 CA Leiden, the Netherlands
\and
Leiden Observatory, Leiden University, PO Box 9513, 2300 RA Leiden, the Netherlands
\and
Lawrence Livermore National Laboratory, Livermore, CA 94550, USA
\and
University of Maryland College Park, Department of Astronomy, College Park, MD 20742, USA
\and
NASA Goddard Space Flight Center (GSFC), Greenbelt, MD 20771, USA
\and
Center for Research and Exploration in Space Science and Technology, NASA GSFC (CRESST II), Greenbelt, MD 20771, USA
\and
Department of Physics, Technion, Haifa 32000, Israel
\and
MIT Kavli Institute for Astrophysics and Space Research, Massachusetts Institute of Technology, Cambridge, MA 02139, USA
\and
ESA European Space Astronomy Centre (ESAC), Camino Bajo del Castillo s/n, 28692 Villanueva de la Cañada, Madrid, Spain
\and
Department of Physics and Astronomy, James Madison University, Harrisonburg, VA 22807, USA
\and
ESA European Space Research and Technology Centre (ESTEC), Keplerlaan 1, 2201 AZ, Noordwĳk, the Netherlands
\and
Department of Physics, University of Tokyo, 7-3-1 Hongo, Bunkyo-ku, Tokyo 113-0033, Japan
\and
Department of Astronomy, University of Michigan, 1085 South University Avenue, Ann Arbor, MI, 48109, USA
\and
Science Research Education Unit, University of Teacher Education Fukuoka, Munakata, Fukuoka 811-4192, Japan
\and
Astronomical Institute, Tohoku University, 6-3 Aramakiazaaoba, Aoba-ku, Sendai, Miyagi 980-8578, Japan
\and
Institute of Space and Astronautical Science (ISAS), Japan Aerospace Exploration Agency (JAXA), Kanagawa 252-5210, Japan
\and
Graduate School of Science and Engineering, Kagoshima University, Kagoshima, 890-8580, Japan
}
\date{Received 22 May 2025 / Accepted 11 June 2025}

\abstract
{
We present our study of the X-Ray Imaging and Spectroscopy Mission (XRISM) observation of the Seyfert-1 galaxy NGC 3783. XRISM’s Resolve microcalorimeter has enabled, for the first time, a detailed characterization of the highly ionized outflows in this active galactic nucleus. Our analysis constrains their outflow and turbulent velocities, along with their ionization parameter ($\xi$) and column density (\NH). The high-resolution Resolve spectrum reveals a distinct series of Fe absorption lines between 6.4 and 7.8 keV, ranging from \ion{Fe}{xviii} to \ion{Fe}{xxvi}. At lower energies (1.8--3.3 keV), absorption features from Si, S, and Ar are also detected. Our spectroscopy and photoionization modeling of the time-averaged Resolve spectrum uncovers six outflow components, five of which exhibit relatively narrow absorption lines with outflow velocities ranging from 560 to 1170 km~s$^{-1}$. In addition, a broad absorption feature is detected, which is consistent with \ion{Fe}{xxvi} outflowing at 14,300 \kms (0.05~$c$). The kinetic luminosity of this component is 0.8--3\% of the bolometric luminosity. Our analysis of the Resolve spectrum shows that more highly ionized absorption lines are intrinsically broader than those of lower-ionization species, indicating that the turbulent velocity of the six outflow components (ranging from 0 to 3500 \kms) increases with $\xi$. Furthermore, we find that the column density (\NH) of the outflows generally declines with the ionization parameter up to $\logxi = 3.2$ but rises beyond this point, suggesting a complex ionization structure. The absorption profile of the \ion{Fe}{xxv} resonance line is intriguingly similar to UV absorption lines (\lya\ and \ion{C}{iv}) observed by the \textit{Hubble} Space Telescope, from which we infer that the outflows are clumpy in nature. Our \xrism/Resolve results from lower- and higher-ionization regimes support a ``hybrid wind'' scenario in which the observed outflows have multiple origins and driving mechanisms. We explore various interpretations of our findings within active galactic nucleus wind models.
}


\authorrunning{M. Mehdipour et al.}

\titlerunning{Delving into the depths of NGC 3783 with XRISM. I.}

\maketitle

\nolinenumbers

\section{Introduction}
\label{sect_intro}

Outflows and winds in active galactic nuclei (AGNs) serve as crucial links between supermassive black holes and their surrounding environments. These outflows carry matter away from the central black hole, dispersing it throughout the host galaxy. This transfer of mass and energy plays a key role in the coevolution of supermassive black holes and their galaxies, influencing feedback mechanisms that affect AGN activity and star formation \citep{King15,Gasp17}. Therefore, understanding the physical properties, energetics, and driving mechanisms of AGN winds is essential for determining their role in AGN feedback and evaluating their impact on the interstellar medium.

The dynamics, kinematics, and ionization structure of ionized outflows, ranging from the vicinity of the accretion disk to the outer regions of the host galaxy, remain poorly understood. This uncertainty makes it difficult to determine how these outflows transfer momentum and energy into the galaxy and influence their surroundings. Ionized outflows (e.g., \citealt{Laha14}) have been observed at different scales, each exhibiting distinct characteristics: the micro (subparsec) scale, associated with the accretion disk and broad-line region (BLR); the meso (parsec) scale, linked to the torus and narrow-line region (NLR); and the macro (kiloparsec) scale, extending into the host galaxy \citep{Laha21,Gall23}. However, the formation mechanisms of these various outflows and their interconnections remain uncertain. Their connection to galactic molecular outflows is not well understood either. Key questions persist regarding the origin of ionized outflows (the disk or torus) and the mechanisms driving them (thermal, radiative, or magnetic). The physical factors that govern the launching and duty cycles of these winds, as well as how wind parameters scale with redshift and AGN properties (e.g., luminosity), remain open questions.

The Resolve microcalorimeter on board the X-ray Imaging and Spectroscopy Mission (XRISM; \citealt{Tash20,Tash25}) offers an exceptional combination of energy resolution (${\sim 4.5}$ eV full width at half maximum, FWHM) and sensitivity at hard X-ray energies (1.8–12 keV), making it invaluable for studying highly ionized outflows that exhibit X-ray absorption features in the Fe K band (6–8 keV). These outflows can have velocities ranging from moderate (a few hundred \kms) to relativistic speeds, i.e. the ultra-fast outflows (UFOs; \citealt{Tomb10}). Studying highly ionized outflows with previous X-ray missions has been challenging due to limitations in both spectral resolution and sensitivity, leading to considerable uncertainties in their kinematics and ionization structure. Consequently, key aspects of the origin, launch mechanisms, and kinetic power of these outflows remain poorly understood. XRISM/Resolve’s high resolving power enables a detailed probing of these highly ionized outflows, which in turn enables a precise determination of their parameters, as first demonstrated for NGC 4151 by \citet{xrism24}.

The bright Seyfert-1 galaxy \ngc, rich in spectral lines, serves as an excellent laboratory for studying ionized outflows in AGNs. It is an ideal target for XRISM because it is exceptionally bright in the Fe K band and has previously exhibited evidence of highly ionized outflows, seen as both narrow \citep{Kasp02} and broad \citep{Mehd17} X-ray absorption features. Additionally, it exhibits a clear and well-defined spectrum of ionized outflows in the soft X-rays \citep{Beh03,Mao19,Gu23,Li25a} and UV \citep{Gab05,Scot14,Kris19}. A 900~ks \chandra High-Energy Transmission Grating (HETG; \citealt{Cani05}) spectrum revealed a 3$\sigma$ detection of a narrow \fexxv resonance line, as well as marginal detections of other transitions \citep{Kasp02,Kron03,Kron05,Yaqo05}. Evidence of \fexxv\ absorption has also been detected in \xmm European Photon Imaging Camera (EPIC; \citealt{Stru01,Turn01}) spectra \citep{Reev04,Cost22}, suggesting that the highly ionized absorption is variable over time. Remarkably, \ngc also undergoes periods of transient obscuration events caused by disk winds, during which new blueshifted and broad absorption features appear in the Fe K band and the UV \textit{Hubble} Space Telescope (HST) Cosmic Origins Spectrograph (COS; \citealt{Gree12}) spectra \citep{Mehd17}.

The \xrism performance verification (PV) observation of \ngc started on July 18, 2024 (04:58 UTC) and ended on July 27, 2024 (16:16 UTC). The total exposure of the Resolve spectrum is 439 ks. Fortunately, at the time of our observation, \ngc was particularly bright in X-rays ($F_{2-10\,{\rm keV}} = 6.0 \times 10^{-11}$~\ergflux). Joint simultaneous observations with \xmm, \nustar, \chandra, NICER, \swift, and HST/COS were also performed. In this first paper, we focus exclusively on modeling the \xrism/Resolve spectrum, without incorporating X-ray data from other telescopes. This approach allows us to fully utilize Resolve's unique capability to measure the properties of highly ionized outflows. Further studies of the outflows, incorporating multi-mission spectral modeling, will be presented in future work. Additionally, studies of emission (reflection) and spectral variability are planned for upcoming papers in our series.

\section{Data reduction and preparation}
\label{sect_data}

\subsection{XRISM data}
\label{sect_xrism}

Our \xrism observation of \ngc (Obs ID: 300050010) was conducted with the gate valve closed and the open filter wheel configuration of Resolve. A brief summary of the data reduction process and systematic errors is provided below, with more detailed descriptions available in our upcoming cross-calibration paper for \ngc \citep{PaperII} as well as in \citet{XRISM25} for PDS 456, which is similar to that of our NGC 3783 observation. We utilized JAXA's pre-pipeline version {\tt 004\_007.20Jun2024\_Build8.012} and pipeline version {\tt 03.00.012.008}. Data analysis was performed using the public XRISM {\tt CALDB version 9} (20240815 release) and the {\tt ftools} package, with additional screening and energy-dependent rise time cuts applied according to \citet{Mochi25} and following the XRISM Quick Start Guide Version 2.1. This CALDB includes the initial updates to the Resolve energy scale and line-spread function files based on in-orbit calibration data. After applying good time interval (GTI) filtering, the total effective exposure time was 439 ks.

We selected only the high-resolution primary (Hp) events for our analysis (see \citealt{Ishi18} for definition of event grading). The count rate was approximately 0.1~s$^{-1}$~pix$^{-1}$ in the central four pixels and ranged from 0.001 to 0.1~s$^{-1}$~pix$^{-1}$ in the outer pixels, where nearly all astrophysical events are expected to be Hp. To mitigate contamination from ``pseudo'' low-resolution secondary (Ls) events and corresponding errors in the normalization of the response matrix file (RMF), all Ls events were excluded before generating the RMF \citep{TTWOF}. The RMF was created using the extra-large (“{\tt X}”) option, which includes all known instrumental effects. Gain tracking was performed using 24 fiducial measurements of the Mn K$\alpha$ line from the $^{55}$Fe filter in the filter wheel. Of the 36 pixels, pixel 12 (calibration pixel) and pixel 27 (which exhibits unpredictable gain jumps) were excluded from the analysis.

The instrumental spectroscopic uncertainties that could impact observational results are uncertainties in the broadband energy scale and the time-dependent gain reconstruction \citep{Ecka24,Port24}, as well as in the energy resolution. These values have been assessed extensively on the ground and in-orbit. The energy scale uncertainty after gain reconstruction is estimated to be 0.34 eV in the 5.4--8.0 keV band for our observation. This value is based on the ${\sim 0.16}$ eV gain reconstruction error at 5.9 keV and the systematic energy scale uncertainty of $\lesssim 0.3$ eV in this band, which are uncorrelated and can be root-sum-squared. The total uncertainty corresponds to 15 \kms at 7 keV. At energies below 5.4 keV, the total energy scale uncertainty is consistent with ${\lesssim 1}$ eV, which corresponds to 150 \kms at 2 keV. The uncertainty on the per-pixel Hp energy resolution values provided in the CALDB is energy-dependent and estimated to be ${\lesssim 0.3}$ eV FWHM from 2--10 keV. Its contribution to the systematic uncertainty is smaller than the statistical error on our best-fit model parameters that we present later in Sect. \ref{sect_model} (Table \ref{table_para}).

The preparation of the background-subtracted Resolve spectrum of \ngc, including the modeling of the non-X-ray background (NXB), as well as the data reduction of Xtend \citep{Noda25}, is further discussed in a forthcoming paper from our campaign \citep{PaperII}, which focuses on the cross-calibration of different instruments. For illustration, Fig.~\ref{fig_nxb} compares the NXB and source spectra, showing that the NXB contribution is minimal in the Resolve spectrum of \ngc.

\subsection{HST data}
\label{sect_hst}

Our new HST/COS observations of \ngc (Program ID: 17273) were conducted on 2024-07-21 over two orbits. The COS data were obtained using the G130M and G160M gratings to capture the \lya and \civ lines. The data were processed with the latest calibration pipeline, {\tt CalCOS v3.6.0}, and the wavelength calibration was verified by checking the observed positions of known Galactic interstellar medium lines. All COS exposures were combined into a single calibrated spectrum, which was then binned by four pixels to improve the signal-to-noise ratio (S/N) while still oversampling the 10-pixel resolution element of the far-UV detector \citep{Fox18}. For further details on COS data preparation, we refer to the previous HST/COS observation of \ngc in \citet{Kris19}.

\begin{figure*}[!tbp]
\centering
\hspace*{-0.0cm}\resizebox{1.0\hsize}{!}{\includegraphics[angle=0]{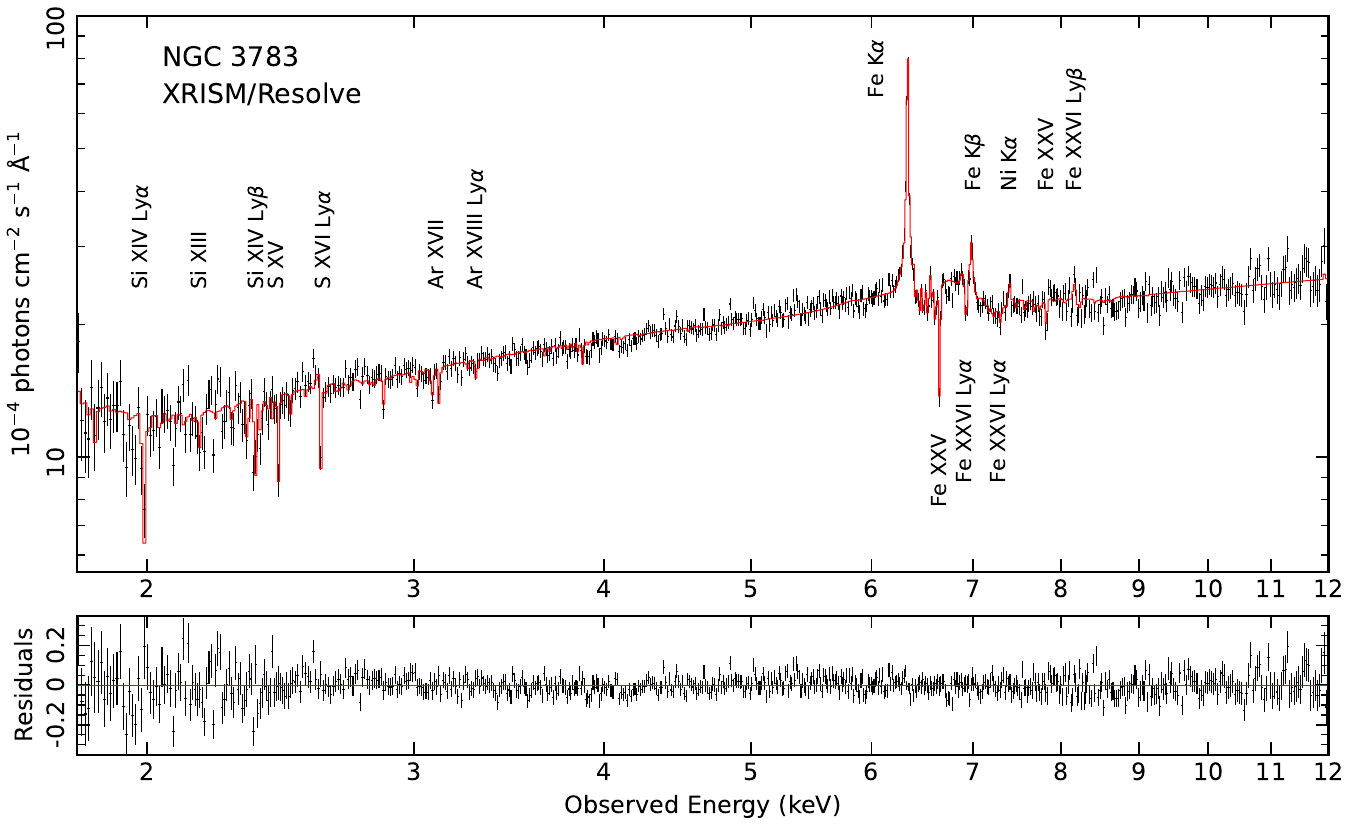}}\vspace{+0.2cm}
\hspace*{-0.0cm}\resizebox{1.0\hsize}{!}{\includegraphics[angle=0]{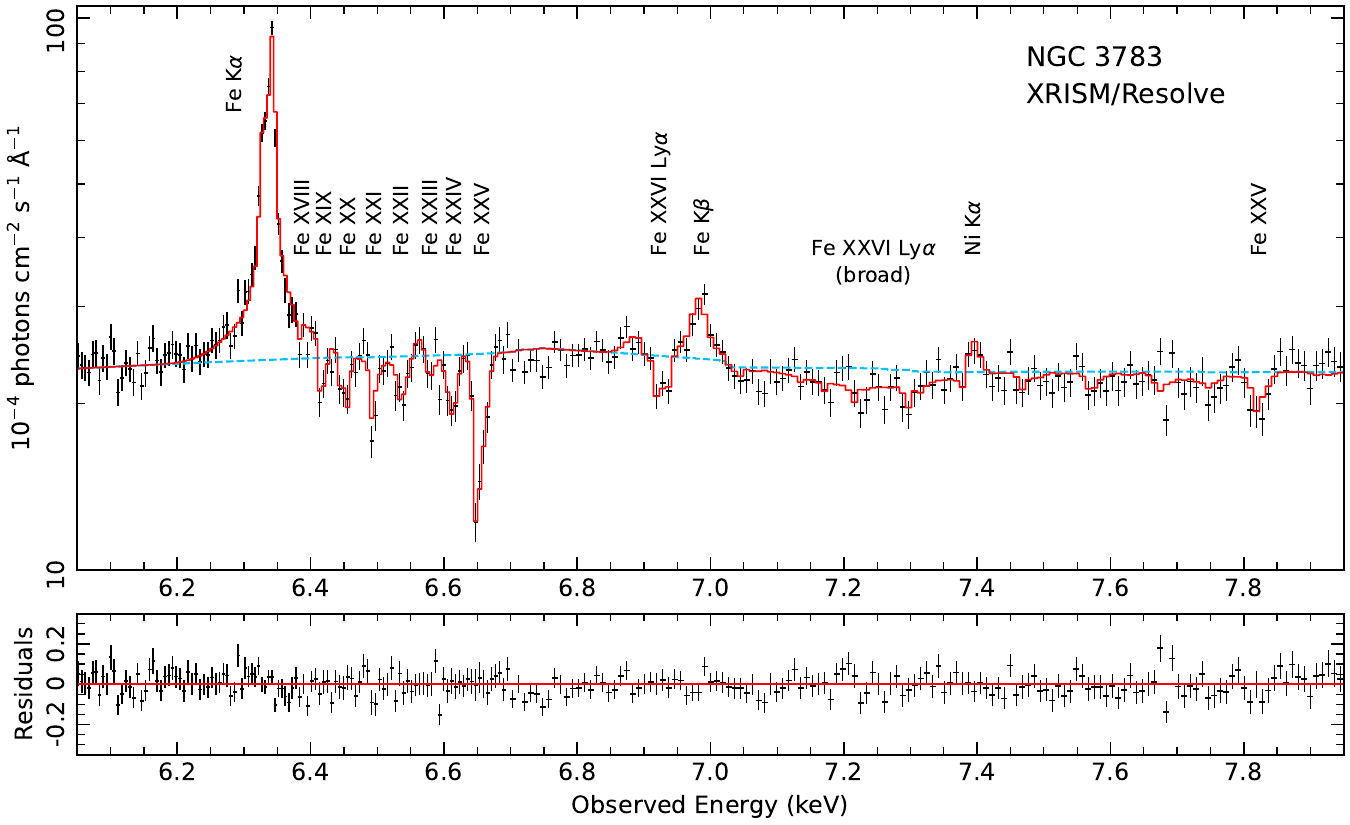}}
\caption{XRISM/Resolve spectrum of \ngc with our best-fit model. The top two panels show the full spectrum and the corresponding fit residuals. The bottom two panels provide a close-up view of the Fe K band and its fit residuals. For clarity of display, the spectrum is additionally binned up. The strongest emission and absorption features are labeled. Our best-fit model (Table \ref{table_para}) is shown in red. The fit residuals are defined as (data $-$ model) / model. For comparison, the dashed blue line in the third panel represents the continuum plus the broad Fe~K emission, excluding any absorption lines.}
\label{fig_spec}
\end{figure*}

\section{Spectral modeling and results}
\label{sect_model}

We modeled the time-averaged \xrism/Resolve spectrum (Fig. \ref{fig_spec}) using \spex\ {\tt v3.08.01} \citep{Kaas96,Kaas24} and its latest atomic database. The {\tt rbin} command of \spex was used to optimally bin both the spectrum and the response file, thereby preventing an oversampling of the data. The underlying theory and algorithms for {\tt rbin}, which account for both the source statistics and the instrumental resolution, are described in the \spex manual (see also \citealt{Kaas16}). The Resolve spectrum was fitted over the 1.8--12~keV band with C-statistics. The cosmological redshift ({\tt reds}) was fixed at 0.009730 \citep{Theu98}, corresponding to a luminosity distance of 41.98 Mpc in \spex based on the cosmological parameters ${H_{0}=70\ \mathrm{km\ s^{-1}\ Mpc^{-1}}}$, $\Omega_{\Lambda}=0.70$, and $\Omega_{\rm m}=0.30$. Galactic X-ray absorption was accounted for using the {\tt hot} model \citep{dePl04,Stee05} in \spex, with the temperature fixed at its minimum value of 0.001~eV, and a column density of $\NH = 9.59 \times 10^{20}$~cm$^{-2}$ \citep{Murp96}. The abundances of all components were set to the protosolar values of \citet{Lod09}.

\subsection{Modeling of the continuum and emission lines}
\label{sect_cont}

The observed continuum was fitted with a simple power-law model ({\tt pow} in \spex). The intrinsic photon index $\Gamma$ of the underlying X-ray power-law continuum is found to be ${\Gamma = 1.79 \pm 0.01}$ from fitting the full Resolve band (1.8--12 keV). This $\Gamma$ is consistent with the \nustar spectrum at higher energies. To model the ``soft X-ray excess,'' we included a warm Comptonization component ({\tt comt}), following the approach used for \ngc in \citet{Mehd17}. However, given that the Resolve spectrum covers only energies above 1.8 keV, the soft excess contribution is minimal (2\% of the flux in the Resolve band). Nonetheless, we included the {\tt comt} component with its parameters fixed to those of the 2001 unobscured model \citep{Mehd17}.

We account for the presence of emission lines in our modeling of the Resolve spectrum. A detailed study of the Fe emission in the Resolve spectrum is presented in an upcoming paper from our campaign by \citet{Li25}. As this study focused on absorption by outflows, we adopted a relatively simple model for the emission lines, ensuring a well-fitted representation of both the continuum and emission features to accurately constrain the absorption lines. Our model includes the following emission features in the Resolve spectrum: \FeKa, \FeKb, Ni~K$\alpha$, \ion{Fe}{xxvi} \lya and \lyb, \ion{Fe}{xxv}, and \ion{S}{xvi} \lya. Each emission line is modeled using a delta function (\delt in \spex) convolved with a Gaussian profile (\vgau), with multiple velocity broadening components applied as needed to achieve a good fit. The centroid of each line is fixed at zero velocity shift in the AGN rest frame, and for doublets, the expected line ratio is applied. The Gaussian velocity broadening of the lines is linked in a physically consistent manner. The velocity widths ($\sigma_v$) of the Gaussian components fitted to the \FeKa emission line range from a few hundred to a few thousand \kms. These are broadly comparable to the widths of the \FeKa components observed in the Resolve spectrum of NGC~4151 \citep{xrism24}. The Resolve spectrum of \ngc also shows the presence of a broad \FeKa component, likely to be relativistic emission from the accretion disk (see \citealt{Li25}). For consistency, we incorporated the relativistic emission component obtained by \citet{Li25} alongside our modeling of the other components using Gaussians. Our tests show that whether the broad \FeKa emission and its associated \FeKb component are modeled with a relativistic or nonrelativistic profile, the absorption modeling results (specifically the number of required components and their parameters) remain essentially unchanged.

\subsection{Photoionized absorption modeling}
\label{sect_photo}

Photoionization calculations were performed using the \pion model \citep{Meh16b,Mill15} in \spex. For this, we adopted the 2001 unobscured spectral energy distribution (SED) of \ngc from \citet{Mehd17}, which is consistent with the intrinsic UV and X-ray continuum of our observation. Using the {\tt xabsinput} program in \spex, which runs the \pion model, we generated tables of ionic concentrations as a function of ionization parameter ($\xi$). These were then used by the \xabs model \citep{Stee03} in \spex to compute the model spectrum. To accurately reproduce all absorption features in the Resolve spectrum (Fig. \ref{fig_spec}), six \xabs components were required. We name these components with a letter in ascending order of outflow velocity \vout, and those with comparable \vout are further sub-labeled by a number in descending order of ionization parameter $\xi$ as shown in Table \ref{table_para}. The ionization parameter ($\xi$), column density (\NH), outflow velocity (\vout), and turbulent velocity (\sigv) of the \xabs components were fitted. We assumed full covering fractions for all \xabs components, as this already provides a good fit to all lines in the Resolve spectrum.

In our model setup, all \xabs components are illuminated by the same ionizing SED and are therefore not treated as sequential layers that shield one another. Since \ngc is unobscured during our observation, there is no strong absorption of the ionizing continuum (in contrast to the obscured 2016 epoch; \citealt{Mehd17}) that would otherwise shield components located farther out. In a layered configuration, the higher-ionization \xabs components would be located closer to the source. However, since these components have relatively low column densities and high ionization parameters, they do not significantly absorb the continuum. As a result, any shielding effect on the ionization state of the outer components would be minimal. Furthermore, as we discuss later in this paper, the outflows in \ngc are found to be clumpy and structurally complex, and thus are unlikely to be arranged in simple sequential layers. Interestingly, study of obscuration in NGC 5548 \citep{Mehd24} has shown that, contrary to the shielding scenario suggested by the appearance of the UV lines, the X-ray absorption lines are unaffected by the obscuration. This indicates that the components do not cover each other in an orderly fashion.

\subsection{Examination of the broad absorption component}
\label{sect_broad}

The Resolve spectrum of \ngc also shows the presence of a relatively broad dip (${\sigv \approx 3500}$ \kms) between 7.1 and 7.4 keV (Fig. \ref{fig_spec}). The equivalent width of this feature is ${17 \pm 3}$\,eV, corresponding to S/N of approximately 6\,$\sigma$. We find that, regardless of how the Fe K emission is modeled (relativistic or nonrelativistic), this broad residual persists below the power-law continuum, which is accurately determined by fitting the full Resolve band (1.8--12 keV) and is consistent with the \nustar spectrum at higher energies. The shape of the residual is consistent with an absorption line rather than an absorption edge. Also, at lower energies, there is no evidence of intrinsic neutral absorption in either the \xrism data or other X-ray observations from our campaign with \xmm and \swift. Furthermore, no evidence of a neutral Fe K edge was previously found in the 900~ks \textit{Chandra}/HETG spectrum of \ngc \citep{Kasp02}. Even during previous epochs when \ngc became obscured \citep{Mehd17}, the obscuring gas was significantly ionized (${\log \xi \sim 1.84}$), with no neutral absorber present. We find that the strength and energy of the residual are inconsistent with either a neutral or ionized absorption edge.

We investigated whether the broad absorption feature could instead arise from a reflection edge, but find this scenario unfeasible. Fits to the \FeKa line using reflection models, such as the {\tt refl} model \citep{Magd95,Zyck99} in \spex and the {\tt MyTorus} table model \citep{Murp09}, show that the associated reflected component is too weak to account for the observed trough. Our tests demonstrate that reflected and scattered emission cannot reproduce the strength or shape of the observed feature. Moreover, our modeling shows that a simple power-law fits the continuum well across the 1.8--12~keV bandpass of Resolve, with no indication of excess emission at higher energies in the spectrum. In our follow-up paper, which presents a detailed analysis of the Fe~K emission in the Resolve spectrum \citep{Li25}, we find that the broad absorption feature persists regardless of the emission model adopted. We therefore conclude that the broad absorption trough is not the result of a reflection edge.

\begin{figure}[!tbp]
\centering
\hspace*{-0.0cm}\resizebox{1.00\hsize}{!}{\includegraphics[angle=0]{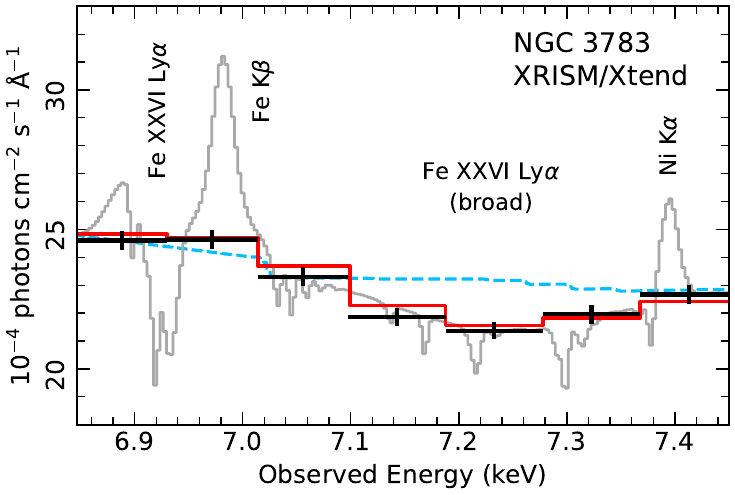}}
\caption{
\xrism/Xtend spectrum of \ngc in the region of the broad \fexxvi absorption feature of Comp.~X. The Xtend data points are shown in black. Our best-fit model to the Resolve spectrum (Fig.~\ref{fig_spec} and Table~\ref{table_para}) is shown in gray. The same model, convolved with the Xtend instrumental response, is overlaid in red. There is close agreement between this convolved model and the Xtend data, demonstrating that our Comp.~X from Resolve is consistent with the Xtend data. The dashed blue line represents the continuum plus the broad Fe~K emission (Sect. \ref{sect_cont}), under which the broad \fexxvi absorption is evident in the Xtend spectrum and is likewise seen in the Resolve spectrum (Fig. \ref{fig_spec}).
}
\label{fig_xtend}
\end{figure}

Adding an additional \xabs component (named Comp. X) enables us to model the residual well as \fexxvi \lya, blueshifted with an outflow velocity of 14,300 \kms. Since this component is associated with a single highly ionized absorption line, its ionization parameter cannot be tightly constrained through spectral fitting. We therefore fixed its ionization parameter to ${\log \xi = 4.0}$ based on our photoionization calculations so that it produces only \fexxvi. Including this component reduces the C-statistic by ${\Delta C = 30}$, further improving the fit. Moreover, we compared this model from Resolve to the Xtend spectrum and find it to be fully consistent. This is demonstrated in Fig.~\ref{fig_xtend}, where the presence of Comp.~X is evident in the Xtend spectrum. Therefore, in our spectral modeling the broad feature is consistent with \fexxvi absorption with a sub-relativistic outflow velocity of 0.05 $c$.

Our best-fit model and residuals are shown in Fig. \ref{fig_spec}, with the corresponding best-fit parameters listed in Table \ref{table_para}. The reduced C-statistic of our best fit is 1.0. The reported statistical uncertainties on the parameters correspond to the 1$\sigma$ confidence level. Figure \ref{fig_trans} illustrates the individual contribution of each \xabs component to the overall absorption. The highest-ionization components (Comps. A1, C, and X) are needed to model \fexxvi absorption at different velocities, while most of \fexxv absorption is produced by Comps. A1 and B. Furthermore, Comp. B is also responsible for absorption by \ion{Fe}{xviii} to \ion{Fe}{xxiv}. Components A2 and A3 reproduce the lower-ionization lines from Si, S, and Ar in the 1.8--3.3 keV band. The C-statistic improvement ($\Delta C$) in fitting the absorption lines by each component is: 41 for Comp. A1, 129 for Comp. A2, 52 for Comp. A3, 168 for Comp. B, 31 for Comp. C, and 30 for Comp. X.

A close-up view of the absorption profile of the \fexxv resonance line ($1s$--$2p$ transition), along with the corresponding best-fit model, is presented in Fig. \ref{fig_profile}. For comparison, the absorption profiles of the \lya and \civ lines from the 2024 HST/COS spectrum are shown in the bottom panel of Fig. \ref{fig_profile}. Furthermore, in Fig. \ref{fig_rel} we examine the relationships between the parameters of the six outflow components. The absorption measure distribution (AMD; \citealt{Holc07}), defined as ${|{\rm d}\,\NH / {\rm d}\, (\log \xi)|}$, is also shown in the top panel of Fig. \ref{fig_rel}. These results are discussed in detail in the following section.

In our \spex modeling, the line energies for the highly ionized \fexxv and \fexxvi transitions are taken from the NIST Atomic Spectra Database \citep{NIST24}, while those for the lower-ionization species are obtained from the Flexible Atomic Code (FAC; \citealt{Gu08}). This implies uncertainties of $\sim$1~eV or better in the line energies, which do not affect the conclusions drawn from the velocities reported in Table \ref{table_para}.

\begin{table}[!tbp]
\begin{minipage}[t]{\hsize}
\setlength{\extrarowheight}{2pt}
\setlength{\tabcolsep}{4pt}
\caption{Best-fit parameters of the outflow components that we have determined from the \xrism Resolve spectrum of \ngc. 
}
\label{table_para}
\centering
\small
\renewcommand{\footnoterule}{}
\begin{tabular}{c | c c c c}
\hline \hline
Comp.  & \vout              & \logxi              & \NH                   & \sigv       \\
       & (\kms)             & (erg~cm~s$^{-1}$)   & ($10^{21}$~\cm)   & (\kms)      \\
\hline
A1      & $590 \pm 30$      & $3.19 \pm 0.07$     & $3.6 \pm 0.8$    & $40 \pm 20$ \\
A2      & $560 \pm 20$      & $2.52 \pm 0.03$     & $11 \pm 1$       & $60 \pm 20$  \\
A3      & $600 \pm 40$      & $1.65 \pm 0.10$     & $23 \pm 2$       & $ < 10$  \\
B       & $700 \pm 40$      & $2.96 \pm 0.07$     & $15 \pm 1$       & $420 \pm 40$  \\
C       & $1170 \pm 100$    & $3.56 \pm 0.10$     & $9 \pm 3$        & $250 \pm 60$   \\
X       & $14{,}300 \pm 1100$  & $4.00$ (f)       & $107 \pm 14$     & $3500 \pm 1000$ \\                                                                                            
\hline
\multicolumn{5}{c}{C-stat / expected C-stat = 3435 / 3437 $\approx$ 1.0} \\
\multicolumn{5}{c}{C-stat / degrees of freedom = 3435 / 3428 $\approx$ 1.0} \\
\hline
\end{tabular}
\end{minipage}
\tablefoot{
Components are labeled with a letter in ascending order of the outflow velocity (\vout), and those with consistent \vout are further sub-labeled by a number in descending order of the ionization parameter ($\xi$). The best-fit parameters of the power-law component (\pow) are $\Gamma = 1.79 \pm 0.01$ with normalization of $4.16 \pm 0.06 \times 10^{51}$ photons~s$^{-1}$~keV$^{-1}$ at 1 keV. 
}
\vspace{0.0cm}
\end{table}

\section{Discussion}
\label{sect_discuss}

\subsection{Physical structure of the highly ionized outflows}
\label{sect_struc}

The \xrism/Resolve spectrum of \ngc has unveiled a rich array of well-resolved absorption lines, providing a detailed probe of the complex structure of the ionized outflows in this AGN. The high spectral resolution of Resolve allows the disentangling of ionization and velocity components in highly ionized outflows, a task previously unattainable. Our spectroscopic and photoionization modeling has identified six distinct absorption components, spanning a wide range of ionization states ($\logxi = 1.65$ to 4.0) and outflow velocities (560 to 14,300 \kms). For the first time, the high-resolution Resolve spectrum allows direct measurements of the turbulent velocity (\sigv) for individual absorption components. The associated photoionized emission lines appear faint in the Resolve spectrum due to both the dominance of numerous absorption lines and the likely small covering factor of the emission region.

The \xrism/Resolve studies of outflows in the quasar PDS~456 \citep{XRISM25}, the Seyfert-1 galaxies NGC~4151 \citep{xrism24}, and \ngc\ (this work) are commonly finding signatures of outflows in the Fe~K band with similar charge states. Interestingly, these results suggest that both the UFOs and the slower highly ionized outflows in AGNs are complex structures consisting of multiple ionization and velocity components. Five of the absorption components in \ngc (Comps. A1 to C) are consistent with ionized outflows (the so-called warm absorbers; \citealt{Yama24}) that are typically associated with the torus and the NLR, while the broader absorption of Comp. X suggests an association with the BLR. Interestingly, the trend shown in Fig. \ref{fig_rel} (middle panel) indicates that \sigv increases with $\xi$, suggesting that regions of higher ionizations are associated with more dynamic and energetic gas movements. This may be because gas closer to the black hole is more highly ionized and turbulent, influenced by faster Keplerian motion in the stronger gravitational potential. Additionally, in magnetohydrodynamic (MHD) models of AGN outflows (see, e.g., \citealt{Fuku15,Fuku22}), such relations between \sigv and $\xi$ can be produced by magnetically driven winds.

Previously, target-of-opportunity observations of \ngc with \xmm and HST/COS revealed that during an obscuration event in 2016, a new broad absorption feature emerged in the Fe K band \citep{Mehd17}. This component exhibited an outflow velocity of 2300 \kms, similar to the broad UV absorption lines (e.g., \civ) associated with the obscurer, suggesting that the obscuring gas spans a wide range of ionization states. However, the broad absorption component detected in the 2024 Resolve spectrum is unrelated to the obscurer, as it exhibits significantly higher outflow velocities. Additionally, our 2024 X-ray observation does not show signs of obscuration. The newly identified Comp. X, which would have been challenging to detect with \xmm, likely originates closer to the black hole than the obscurer.

\begin{figure}[!tbp]
\centering
\hspace*{-0.0cm}\resizebox{1.00\hsize}{!}{\includegraphics[angle=0]{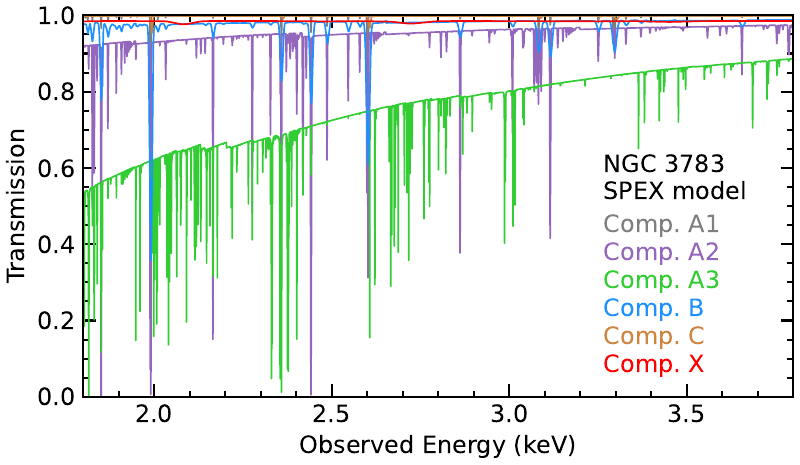}}\vspace{-0.0cm}
\hspace*{-0.0cm}\resizebox{1.00\hsize}{!}{\includegraphics[angle=0]{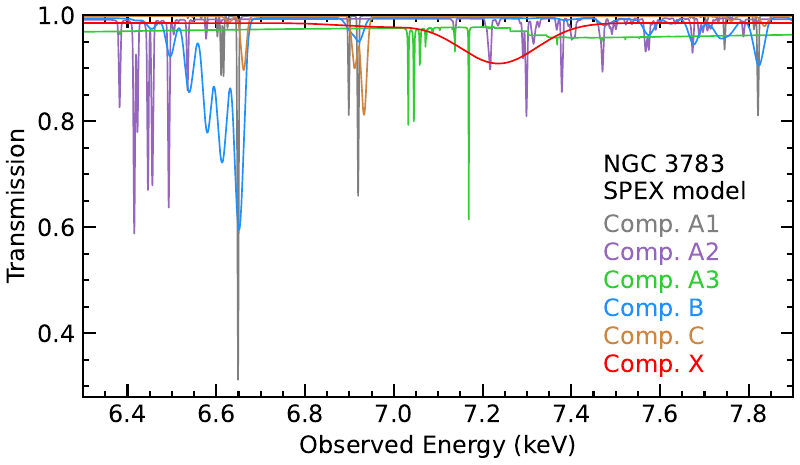}}
\caption{Transmission model of the six outflow components. The spectra correspond to the best-fit model parameters of Table \ref{table_para}. Components are labeled with a letter in increasing order of the outflow velocity (\vout), and those with similar \vout are further sub-labeled by a number in decreasing order of the ionization parameter ($\xi$).}
\label{fig_trans}
\end{figure}

\begin{figure}[!tbp]
\centering
\hspace*{-0.3cm}\resizebox{0.92\hsize}{!}{\includegraphics[angle=0]{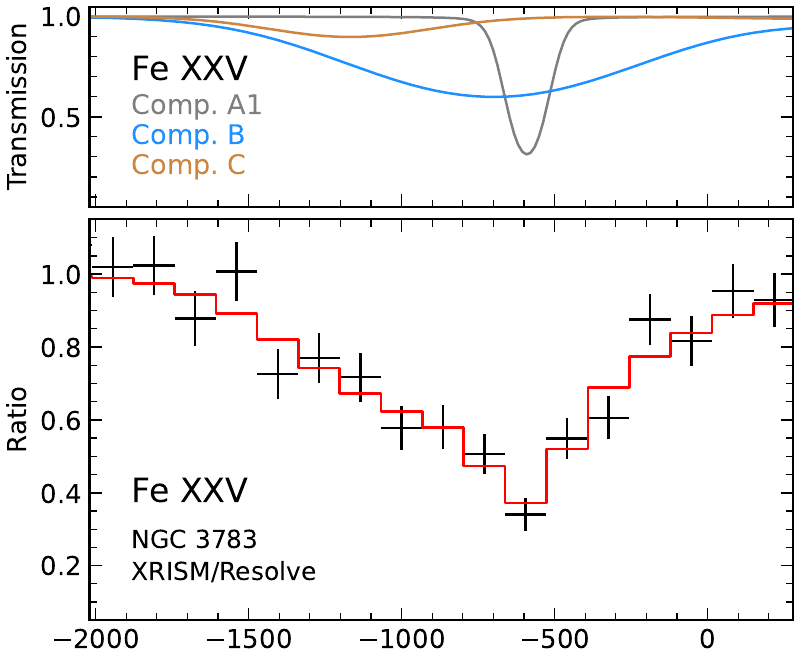}}\vspace{-0.0cm}
\hspace*{-0.3cm}\resizebox{0.92\hsize}{!}{\includegraphics[angle=0]{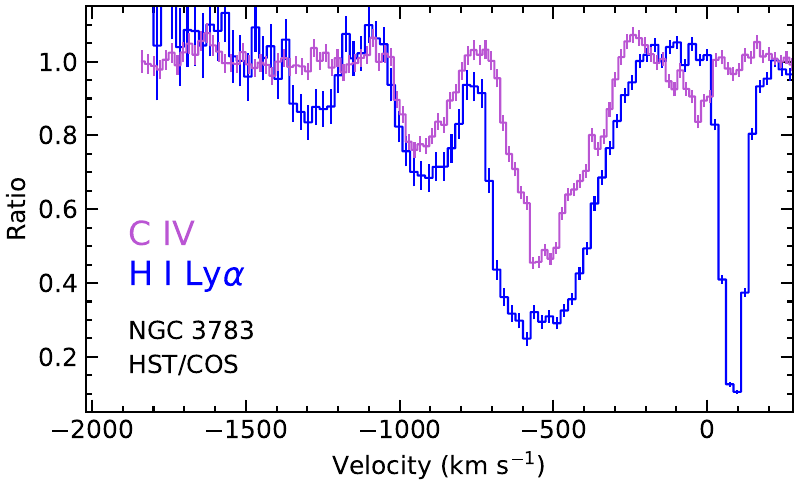}}
\caption{Absorption profile of the \ion{Fe}{xxv} resonance line in the XRISM Resolve spectrum compared to those of \lya and \civ in the 2024 HST COS spectrum. The data are normalized to the continuum, showing the flux ratio on the y-axis. The red model in the middle panel corresponds to the best-fit model shown in Fig. \ref{fig_spec}, with its parameters given in Table \ref{table_para}. The top panel illustrates how the individual components of the model contribute to the \fexxv absorption. In this figure negative velocity corresponds to an outflow, while positive corresponds to an inflow.}
\label{fig_profile}
\end{figure}

The comparison of absorption line profiles for \fexxv, \civ, and \lya in Fig. \ref{fig_profile} provides valuable insights and offers a model-independent way to examine their shapes. While these lines show both similarities and differences, they all exhibit absorption across a range of velocities, indicating that the outflowing gas consists of multiple velocity components. The \fexxv profile appears asymmetric, with absorption extending to higher velocities. We note that the broadening of the \fexxv\ absorption in Fig. \ref{fig_profile} is almost entirely due to the resonance line in each of the three \xabs components, with the contribution from the intercombination line being minimal. Nonetheless, the spectral fitting accounts for all transitions, allowing us to reliably determine the broadening of each component.

The broad and asymmetric absorption profiles observed in \fexxvi (Figs.~\ref{fig_spec} and \ref{fig_xtend}) and \fexxv (Fig.~\ref{fig_profile}) resemble the MHD wind absorption profiles predicted by the simulations of \citet{Fuku22}, which exhibit comparable absorption troughs. Interestingly, there is also a striking resemblance between the \fexxv and UV lines (Fig.~\ref{fig_profile}), as all three exhibit their deepest absorption trough at the same velocity ($\sim-600$~\kms). There is an interplay between ionization and velocity in shaping the observed absorption features. Figure \ref{fig_profile} (top panel) shows that the highly ionized Comps. B and C of the outflows are intrinsically broader (due to higher turbulence) than their UV counterparts. This intrinsic line broadening results in blending of the spectral components. We note that while we attribute the intrinsic line broadening of each component to a single turbulence parameter, in reality each component may have an even more complex velocity structure, which would result in ``velocity shear.'' However, in practice it is challenging to constrain such details from spectral fitting alone. Nonetheless, by modeling the full \xrism/Resolve spectrum, the individual outflow components can be disentangled and parameterized, as shown in Fig.~\ref{fig_trans}, the top panel of Fig.~\ref{fig_profile}, and Table \ref{table_para}.

The bottom panel of Fig.~\ref{fig_profile} shows that \lya traces the greatest number of absorption components. Notably, the highest velocity component at approximately –1300 \kms and the lowest at +100 \kms are observed in \lya. The +100 \kms \lya component likely originates from distant neutral or weakly ionized gas in the AGN's host galaxy, slowly inflowing toward the nucleus. Such a neutral or low-ionization gas would not be expected to have a corresponding high-ionization counterpart, explaining the absence of absorption in the \fexxv profile at this velocity (Fig. \ref{fig_profile}, middle panel). Interestingly, the –1300 \kms \lya component appears to correspond to Comp. C in X-rays, whose \fexxv absorption model is shown in the top panel of Fig. \ref{fig_profile}. Neutral hydrogen can persist even in highly ionized gas due to its high abundance and the inhomogeneous nature of the medium. For example, \lya counterparts of X-ray UFOs have been identified in other AGNs \citep{Kris18a,Mehd22b}. As shown in the bottom panel of Fig. \ref{fig_profile}, the \lya outflow components are generally broader than those of \civ, suggesting that \lya traces gas that is more turbulent and closer to the black hole. Additionally, both \fexxv and \lya profiles extend to higher velocities than \civ, indicating that more highly ionized gas is associated with faster outflows.

In \ngc, we find gas spanning a wide range of ionization states, with similar outflow velocities detected in both X-ray and UV spectra (Fig. \ref{fig_profile}). This indicates a clumpy outflow structure, in which denser (cooler) clouds are embedded within a more diffuse (hotter) medium. Such clumpy outflows are commonly seen in joint X-ray and UV spectroscopic studies of AGNs (e.g., \citealt{Mehd22c,Zaid24}). Various thermal and hydrodynamic instabilities have been proposed to explain the formation of clumpy outflows in AGNs (e.g., \citealt{Take13,Dann20,Wate22}). Alternatively, disk wind models (e.g., \citealt{Fuku17,Fuku24}) have been shown to reproduce the spectral characteristics of a wide range of AGN outflows, from UFOs to broad absorption (obscurers) and narrow absorption (warm absorber) outflows.

\subsection{Energetics and origin of the highly ionized outflows}
\label{sect_energ}

We derived estimates for the kinetic luminosity of the outflows, ${L_{\rm kin} = 1/2 \, \dot{M}_{\rm out} \, v_{\rm out}^2 = 1/2 \, \mu \, m_{\rm p} \, N_{\rm H} \, R \, \Omega \, C_{\rm V} \, v_{\rm out}^3}$, where $\dot{M}_{\rm out}$ is the mass outflow rate, $\mu$ the mean atomic weight per proton (${\approx 1.43}$, from \spex), $m_{\rm p}$ the proton mass, $\Omega$ the solid angle, $C_{\rm V}$ the volume filling factor, and $R$ the radial distance from the source. For $R$, we assumed minimum and maximum constraints following \citet{Blu05}. The minimum was set by the escape velocity condition: ${R \ge 2 \, G \, M_{\rm BH} / v_{\rm out}^2}$, where the black hole mass ${M_{\rm BH} = 2.82 \times 10^7}$~$M_{\odot}$ \citep{Bent21}. The maximum is based on a thin-shell scenario: ${R \le L_{\rm ion} \, C_{\rm V} / (\xi \, N_{\rm H})}$, where the 1--1000 Ryd ionizing luminosity ${L_{\rm ion} = 6.4 \times 10^{43}}$~erg~s$^{-1}$ \citep{Mehd17} is from the SED described in Sect.~\ref{sect_photo}. We adopted a time-averaged bolometric luminosity of ${L_{\rm bol} = 2.1 \times 10^{44}}$~erg~s$^{-1}$ from the broadband continuum modeling of \citet{Mehd17}. This corresponds to an Eddington luminosity ratio of ${L_{\rm bol} / L_{\rm Edd} = 0.06}$. Using a fiducial value of ${\Omega \sim 2\pi}$ \citep{Crens12}, and a volume filling factor ${C_{\rm V} \sim 0.2}$ \citep{XRISM25} from similarly clumpy, highly ionized outflows in PDS~456, we find that for Comp.~X, $L_{\rm kin} / L_{\rm bol}$ ranges from 0.008 to 0.03. This suggests that Comp.~X can contribute significantly to AGN feedback, which requires $L_{\rm kin} / L_{\rm bol} \gtrsim 0.005$ \citep{Hopk10}. Using the minimum and maximum values of $R$ for Comp.~X ($0.4$--$1.2 \times 10^{16}$~cm), its inferred density ${n_{\rm H} \sim N_{\rm H} / (C_{\rm V}\, R) \sim 0.4}$--$1.5 \times 10^{8}$~cm$^{-3}$ is interestingly similar to that of the clumps in the UFOs of PDS~456 (${n_{\rm H} \sim N_{\rm H} / d_{\rm clump} \sim 0.8}$--$6.7 \times 10^{8}$~cm$^{-3}$; \citealt{XRISM25}). For the other slower components in \ngc (Comps. A to C), $L_{\rm kin}$ is much smaller, with a combined $L_{\rm kin} / L_{\rm bol}$ of 0.0002--0.0006. This means that, even assuming these slower components are escaping winds (with some possibly being failed winds), their kinetic luminosity is nonetheless insignificant in either scenario.

\begin{figure}[!tbp]
\centering
\hspace*{-0.05cm}\resizebox{0.917\hsize}{!}{\includegraphics[angle=0]{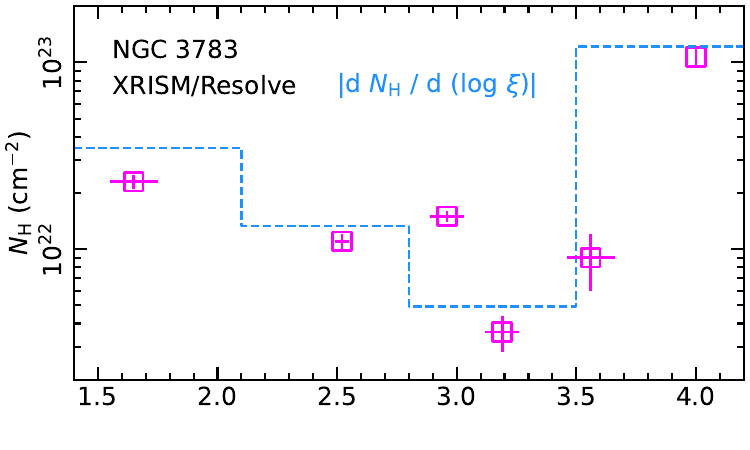}}\vspace{-0.4cm}
\hspace*{-0.0cm}\resizebox{0.92\hsize}{!}{\includegraphics[angle=0]{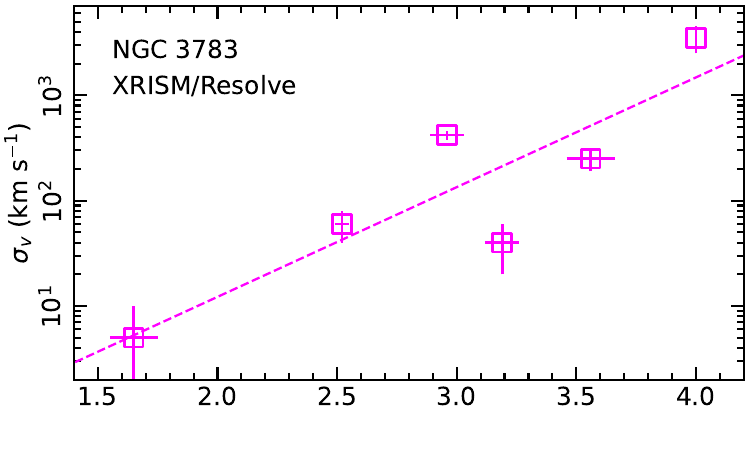}}\vspace{-0.4cm}
\hspace*{-0.0cm}\resizebox{0.92\hsize}{!}{\includegraphics[angle=0]{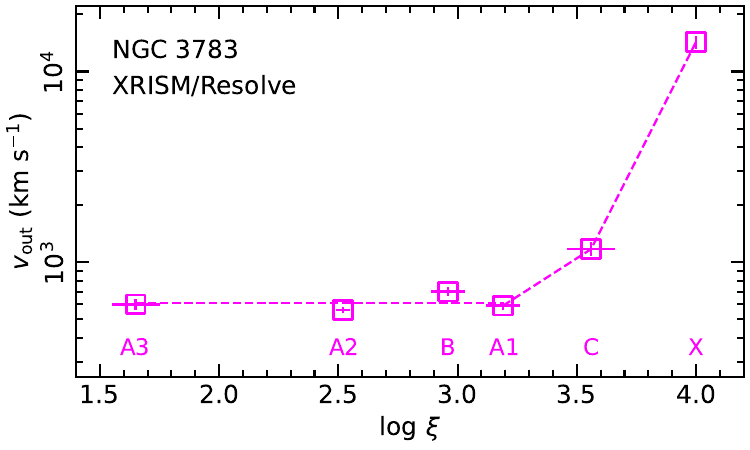}}
\vspace{-0.1cm}
\caption{Relations between the parameters of the six outflow components derived from the \xrism Resolve spectrum of \ngc. The top panel displays the column density (\NH) and the AMD as functions of the ionization parameter (\logxi). The middle and bottom panels show the turbulent velocity (\sigv) and outflow velocity (\vout) as functions of \logxi, respectively. The component label (Table \ref{table_para}) for each data point is shown along the lower edge of the bottom panel.}
\label{fig_rel}
\end{figure}

The relations in Fig. \ref{fig_rel} provide further insight into the properties of outflows observed with \xrism. The AMD is a powerful tool for probing the multiphase nature of AGN outflows \citep{Holc07,Beh09,Ster14,Adhi19}. The radial density of the outflow, as inferred from the AMD, is useful for linking observations to physical models of AGN wind launching and driving mechanisms. The AMD derived from \xrism/Resolve (Fig. \ref{fig_rel}, top panel) offers significantly greater detail, especially at higher ionization parameters, compared to those obtained with previous X-ray missions. Our results reveal that \NH and the AMD generally decline from ${\logxi = 1.65}$ to 3.2, before rising at higher ionization parameters (Fig.~\ref{fig_rel}, top panel). While a roughly flat AMD shape in the lower-ionization regime ($\logxi < 3.2$) cannot be conclusively ruled out, nonetheless, our interpretation of the AMD (discussed below) remains unchanged regardless of whether the shape is declining or flat. Additionally, we emphasize that the significant increase in \NH due to Comp. X is robust, despite its ionization parameter being fixed to the minimum feasible $\xi$ in our modeling, as a higher $\xi$ would require an even greater \NH. Interestingly, while the faster outflow components generally exhibit higher turbulent velocities (Table~\ref{table_para}), a significant jump in the outflow velocity is observed above $\logxi = 3.2$ (Fig.~\ref{fig_rel}, bottom panel). These AMD and velocity trends are not easily attributable to a single wind-driving mechanism, pointing instead to a more complex outflow structure in \ngc. 

Our results point to the presence of a ``hybrid wind'' that gives rise to two distinct patterns in the lower-ionization and higher-ionization regimes. The declining AMD trend at lower ionizations ($\logxi < 3.2$) is consistent with thermally driven winds, as shown by \citet{Dyda17}, while the increasing trend at higher ionizations ($\logxi > 3.2$) aligns with expectations for magnetically driven winds (e.g., \citealt{Fuku15}). Moreover, the significant increase in outflow velocity at higher ionizations further supports this interpretation, in which the slower outflows are thermal winds (e.g., \citealt{Mizu19}) and the faster outflows are magnetic winds (e.g., \citealt{Fuku10}). Thermally driven winds are inherently slow, typically outflowing at hundreds of \kms (e.g., \citealt{Gang21}), consistent with our Comps. A1, A2, A3, and B. In contrast, magnetically driven winds can reach velocities of several thousand \kms, and in some cases, reach relativistic speeds, in line with our Comps. C and X. Therefore, our findings suggest that the absorption along our line of sight originates from two distinct regions: a highly ionized, magnetically driven disk wind, surrounded by thermally driven ionized outflows farther out. We note that radiatively driven outflows (e.g., \citealt{Prog00,Gius19,Wate21,Mizu21}) can also contribute significantly in the hybrid wind scenario that we observe in \ngc. In this Paper~I of our series, we focus on the kinematic and ionization structure of the highly ionized outflows in the time-averaged \xrism/Resolve spectrum. To gain a more complete understanding, all phases of the outflows, including those in the soft X-ray (\xmm/RGS) and UV (HST/COS), must be studied alongside the \xrism/Resolve results. In our follow-up papers, we plan to carry out such a multiwavelength investigation.

\section{Conclusions}
\label{sect_concl}

Our XRISM/Resolve analysis of NGC 3783 reveals multicomponent, highly ionized outflows in this AGN. We identify six absorption components: five with relatively narrow absorption lines and moderate outflow velocities (560–1170 km~s$^{-1}$), and one broad absorption component outflowing at sub-relativistic speeds (0.05\,$c$). With a kinetic luminosity of 0.8--3\% of the bolometric luminosity, this sub-relativistic component is likely energetically significant for AGN feedback. The Resolve spectrum shows that higher-ionization absorption lines (such as \ion{Fe}{xxvi} and \ion{Fe}{xxv}) are generally broader than those of lower-ionization species, suggesting that the gas closer to the black hole is more highly ionized and more turbulent, likely due to increased Keplerian motion in the deeper gravitational potential. Additionally, the column density (\NH) generally decreases with the ionization parameter from \logxi = 1.65 to 3.2 but significantly increases at higher values, hinting at a complex ionization structure along our line of sight. A comparison of the kinematics of the highly ionized outflows with the UV absorption lines (\lya and \ion{C}{iv}) observed by HST/COS reveals that gas across a wide range of ionization states outflows at similar velocities, which is indicative of a clumpy, multi-zone outflow structure. The trends observed by \xrism/Resolve in the parameters of the outflow components in the lower- and higher-ionization regimes suggest a hybrid wind scenario in which a faster, magnetically driven disk wind is surrounded by slower, thermally driven outflows that extend farther out. These findings demonstrate the power of high-resolution X-ray spectroscopy in the Fe K band for probing AGN winds. When combined with multiwavelength observations of the outflows, they provide crucial insights into the structure and driving mechanisms of AGN outflows.

\begin{acknowledgements}
M. Mehdipour acknowledges support from NASA XRISM Guest Scientist (XGS) grant (80NSSC23K0995). This work is also supported by NASA through a grant for HST program number 17273 from the Space Telescope Science Institute, which is operated by the Association of Universities for Research in Astronomy, Incorporated, under NASA contract NAS5-26555. Support from NASA NuSTAR grant 80NSSC25K7126 is acknowledged. SRON is supported financially by NWO, the Netherlands Organization for Scientific Research. Part of this work was performed under the auspices of the U.S. Department of Energy by Lawrence Livermore National Laboratory under Contract DE-AC52-07NA27344. K. Fukumura acknowledges support from NASA XGS grant (80NSSC23K1021). This work was supported by JSPS KAKENHI Grant Number JP21K13958. M. Mizumoto acknowledges support from Yamada Science Foundation. We thank G. Kriss, C. Done, L. Gallo, M. Leutenegger, and R. Mushotzky for valuable discussions and helpful feedback. The anonymous referee is acknowledged for their constructive comments and suggestions, which improved the paper.

\end{acknowledgements}

\bibliographystyle{aa}
\bibliography{references}

\begin{appendix}

\section{Non-X-ray background in the Resolve spectrum of NGC 3783}
\label{sect_app}

\begin{figure}[!tbp]
\centering
\hspace*{-0.0cm}\resizebox{1.00\hsize}{!}{\includegraphics[angle=0]{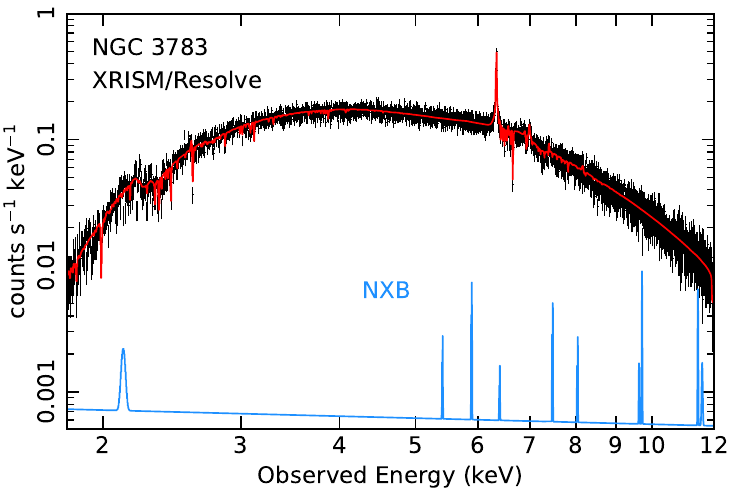}}
\caption{
Comparison of the count rate spectra of the source (black) and the NXB (blue) in the \xrism/Resolve observation of \ngc. The best-fit model, described in Sect.~\ref{sect_model}, is shown in red. The NXB contribution across the bandpass is mostly minimal.
}
\label{fig_nxb}
\end{figure}

We show in Fig.~\ref{fig_nxb} the \xrism/Resolve count rate spectrum of \ngc, together with the corresponding NXB model. The NXB contribution is mostly minimal, becoming marginally relevant toward the ends of the bandpass. Additional details on the Resolve data reduction and NXB modeling are provided in Sect.~\ref{sect_xrism} and \citet{PaperII}.

\end{appendix}

\end{document}